\documentclass[a4paper]{iopart}
\usepackage{graphicx,bm,amssymb,epsfig,cite,hyperref}
\newcommand{\be}{\begin{equation}}
\newcommand{\ee}{\end{equation}}
\newcommand{\bea}{\begin{eqnarray}}
\newcommand{\eea}{\end{eqnarray}}

\def\2t#1#2{\langle\tau_{#1}\tau_{#2}\rangle}

\def\i{{\rm i\,}}

\def\sstyle{\scriptstyle}

\begin{document}

\title{The Razumov-Stroganov conjecture: Stochastic processes, loops and combinatorics}
\author{Jan de Gier}
\address{ARC Centre of Excellence for Mathematics and Statistics of
  Complex Systems, Department of Mathematics and Statistics, The University of
  Melbourne, 3010 VIC, Australia}
  
\maketitle

\subsection*{Introduction}

A fascinating conjectural connection between statistical mechanics and combinatorics has in the past five years led to the publication of a number of papers in various areas, including stochastic processes, solvable lattice models and supersymmetry. This connection, known as the Razumov-Stroganov conjecture, expresses eigenstates of physical systems in terms of objects known from combinatorics, which is the mathematical theory of counting. This note intends to explain this connection in light of the recent papers by Zinn-Justin \cite{ZJ07} and Di Francesco \cite{DF07}.
 
\subsection*{A non-local stochastic process}  
\begin{figure}[h]
\centerline{\includegraphics[width=0.5\textwidth]{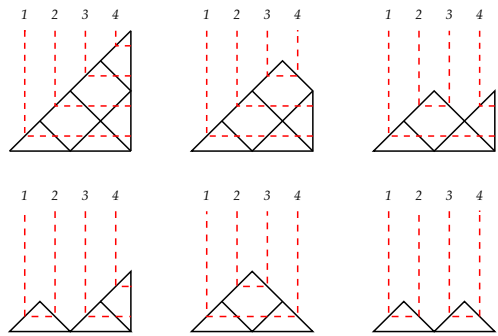}}
\caption{Tile configurations for $L=4$.}
\label{fig:states}
\end{figure}

\noindent
Consider the following stochastic process of stacking and removing tiles and half-tiles in a corner. Allowed configurations of the process are those depicted in Figure~\ref{fig:states} for $L=4$, ignoring for the moment the dashed red lines. 

\begin{figure}[h]
\centerline{
\begin{picture}(320,60)
\put(0,0){\includegraphics[width=130pt]{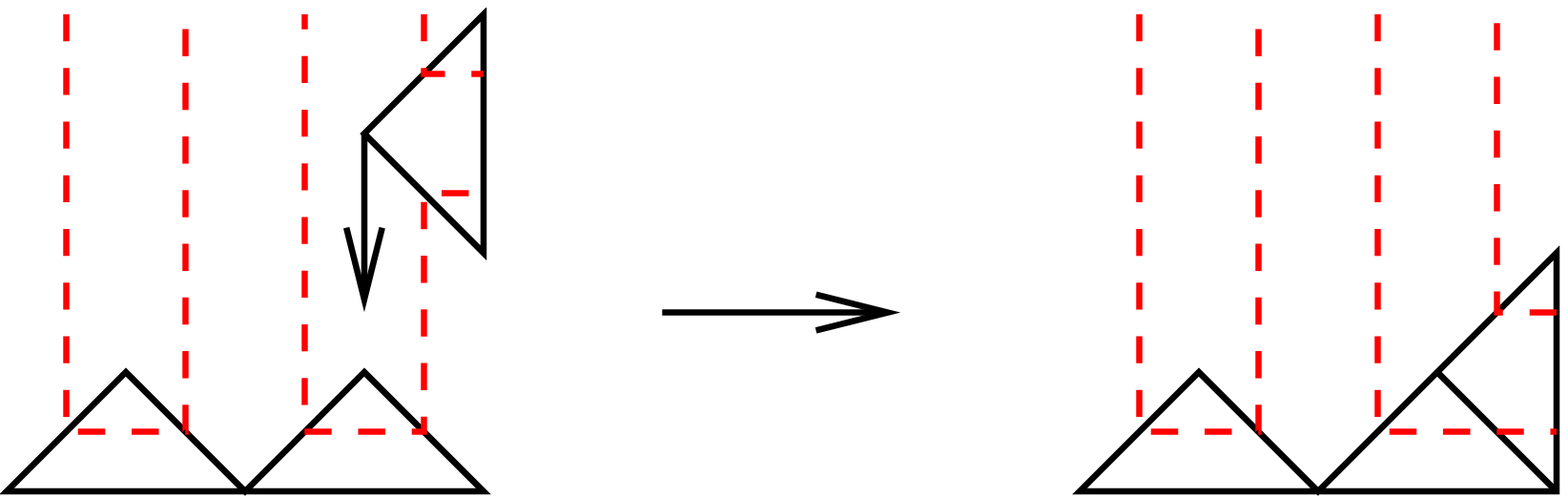}}
\put(190,0){\includegraphics[width=130pt]{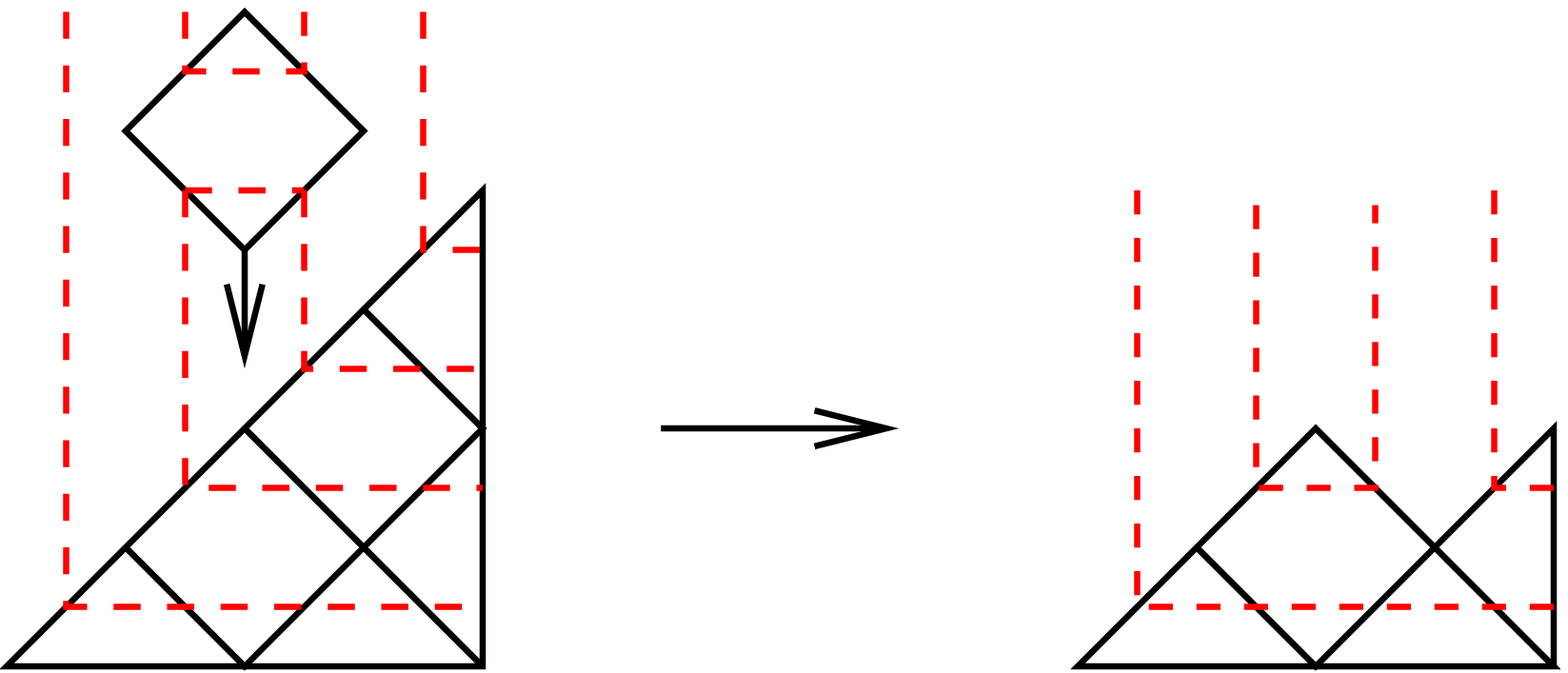}}
\end{picture}}
\caption{Adsorption (Raise) and Desorption (Peel).}
\label{fig:rates}
\end{figure}

\noindent
The transition rates between these configurations, i.e. the rates for stacking and removing tiles, are obtained as follows. When a (half) tile is dropped from above, we join up the dashed red lines, remove any closed loops and identify the resulting picture of dashed red lines with one of the pictures in Figure~\ref{fig:states}. Two examples of this process are given in Figure~\ref{fig:rates}. It easy to convince oneself that when a tile is dropped into a local minimum, it will be adsorbed and the local height will be raised. On the other hand, if a tile is dropped onto a slope in the local height profile, a strip of tiles is peeled off. This raise and peel (RP) process therefore models a system of local growth with non-local macroscopic desorption \cite{GNPR}.  

\noindent
Denote dropping a tile between sites $i$ and $i+1$ by an operator $e_i$, then the transition matrix $H$ of this raise and peel (RP) process is given by
\be
H = \sum_{i=1}^{L-1} (1-e_i) + a (1-e_L) ,
\label{eq:H_alg}
\ee 
where $a$ can be any positive real number. At large times, the RP process converges to a non-equilibrium stationary state $\Psi_L$ given by $H \Psi_L =0$. This stationary state, and the spectral properties of the RP process are very interesting as the process has a dynamic exponent $z=1$ and is an example of a non-equilibrium system whose spectrum is determined by conformal field theory (CFT). 

For what follows it is important to do the simple excercise of writing out the Hamiltonian (\ref{eq:H_alg}) explicitly. For $L=4$, using the pictures in Figure~\ref{fig:states} as basis, we find,
\be
H=-\left(\begin{array}{@{}cccccc@{}}
-3 & a & 0 & 0 & 0 & 0 \\
1 & -2-a & 1 & 0 & 0 & 0 \\
1 & 1 & -2 & 1 & a & 0 \\
1 & 0 & 1 & -2 & 0 & a \\
0 & 0 & 0 & 0 & -2-a & 1 \\
0 & 1 & 0 & 1 & 2 & -1-a
\end{array}\right),
\ee
and it is easy to compute the stationary probability distribution which is given by
\be
\Psi_4=\frac{1}{Z_4}(a^2,3a,2a(3+a),3a(2+a),3,3(2+a)),
\label{eq:psi4}
\ee 
where
\be
Z_4 = 6 a^2 + 18a + 9,
\label{eq:norm4}
\ee
so that the sum over the components of $\Psi_4$ is normalised to one.
\subsection*{Fully packed loops}

Consider now drawing loops on the edges of following grid:

\begin{figure}[h]
\centerline{\includegraphics[width=50pt]{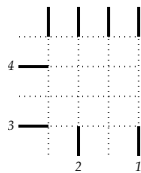}}
\caption{A grid for $L=4$.}
\end{figure}

\begin{figure}[h]
\centerline{
\begin{picture}(320,300)
\put(0,0){\includegraphics[width=320pt]{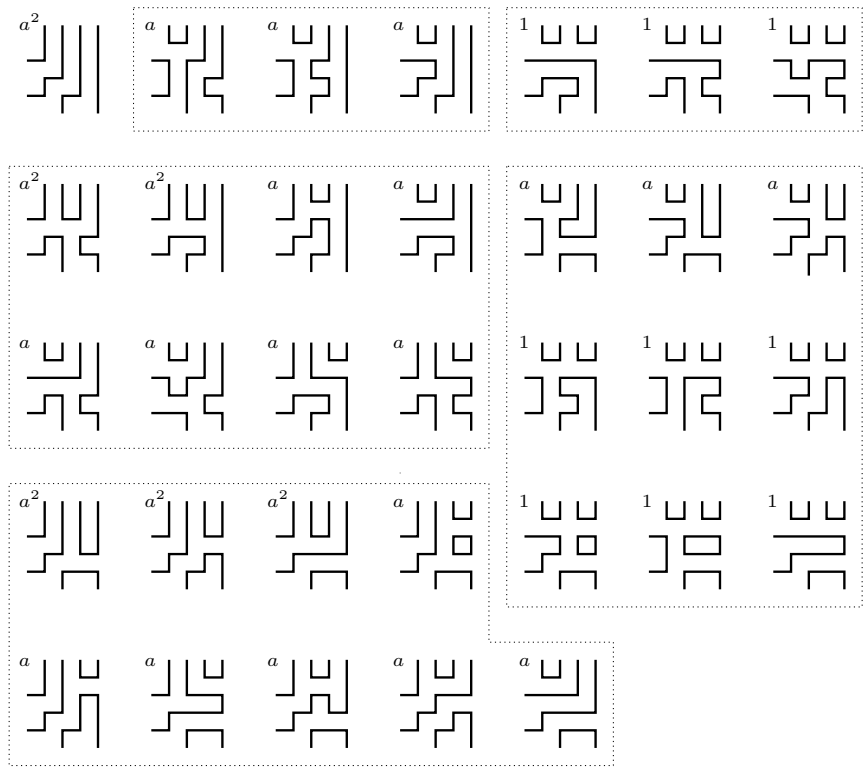}}
\put(4,278){$\sstyle a^2$}\put(51,278){$\sstyle a$}\put(97,278){$\sstyle a$}\put(144,278){$\sstyle a$}\put(191,278){$\sstyle 1$}\put(237,278){$\sstyle 1$}\put(284,278){$\sstyle 1$}
\put(4,218){$\sstyle a^2$}\put(51,218){$\sstyle a^2$}\put(97,218){$\sstyle a$}\put(144,218){$\sstyle a$}\put(191,218){$\sstyle a$}\put(237,218){$\sstyle a$}\put(284,218){$\sstyle a$}
\put(4,158){$\sstyle a$}\put(51,158){$\sstyle a$}\put(97,158){$\sstyle a$}\put(144,158){$\sstyle a$}\put(191,158){$\sstyle 1$}\put(237,158){$\sstyle 1$}\put(284,158){$\sstyle 1$}
\put(4,98){$\sstyle a^2$}\put(51,98){$\sstyle a^2$}\put(97,98){$\sstyle a^2$}\put(144,98){$\sstyle a$}\put(191,98){$\sstyle 1$}\put(237,98){$\sstyle 1$}\put(284,98){$\sstyle 1$}
\put(4,38){$\sstyle a$}\put(51,38){$\sstyle a$}\put(97,38){$\sstyle a$}\put(144,38){$\sstyle a$}\put(191,38){$\sstyle a$}
\end{picture}}
\caption{Weighted FPL diagrams for $L=4$. A diagram with $2j$ vertical
  line segments in the top row is assigned a weight
  $a^j$. Diagrams with the same connectivity are grouped together.}
\label{fig:gsL4W}
\end{figure}

\noindent
The loops have to start and end at the already drawn bonds, they have to visit each site, and are not allowed to cross or overlap. We furthermore assign a weight to these fully packed loop (FPL) configurations in the following way. If an FPL diagram contains $2j$ vertical loop lines in the top row, assign a weight $a^j$ to that particular diagram. The list of weighted diagrams is given in Figure~\ref{fig:gsL4W}. 

\subsection*{Razumov-Stroganov conjectures}
The following should come as a total surprise. 
\begin{enumerate}
\item The weighted sum over the loop diagrams in
  Figure~\ref{fig:gsL4W} is precisely equal to the normalisation of
  $\Psi_4$ given in (\ref{eq:norm4}). This observation generalizes
  into the conjecture that a similar correspondence holds for arbitrary values of $L$.
\item Moreover, \textit{each} element of $\Psi_4$ corresponds to a
  specific set of FPL diagrams, namely precisely those for which the
  way external loop lines on the left and bottom sides of the grid are
  connected is the same as that of the dashed red lines in
  Figure~\ref{fig:states}. The FPL in diagrams in
  Figure~\ref{fig:gsL4W} with the same such connectivities are grouped
  together. Again, this generalizes to arbitrary $L$.
\end{enumerate}

The proof of (i) is the content of a paper by Zinn-Justin \cite{ZJ07}, which is one in a series of recent papers by Di Francesco and Zinn-Justin (see e.g. \cite{DFZJ05}) attempting to prove the so-called Razumov-Stroganov (R-S) conjectures. These conjectures, of which (ii) is a particular instance, relate eigenvectors of certain Hamiltonians to the combinatorics of FPL diagrams (or alternating sign matrices \cite{Bress99} with which FPL diagrams are in simple bijection). A brief history of the R-S conjectures will be given below. 

The paper by Di Francesco \cite{DF07} deals with a similar topic but places more emphasis on the combinatorial aspects surrounding the R-S conjectures. An important ingredient in the papers of Di Francesco and Zinn-Justin is the fact that the eigenvalue equation $H\Psi_L=0$ is equivalent to the $q$-Kniznhik-Zamolodchikov ($q$KZ) equation when $q=\exp(2\i\pi/3)$ \cite{DFZJ05,Pasq05}. The work of Di Francesco however shows that intruiging connections to combinatorics remain when considering the $q$KZ equation for general values of $q$. 

\subsection*{Spin chain}
The origin of the R-S conjectures lies in the one-dimensional quantum spin-1/2 Heisenberg chain, arguably one of the most studied quantum systems. The spin chain is defined by its energy operator which is given by the following Hamiltonian, 
\be
H = -\frac12 \sum_{i=1}^{L-1} J_x \sigma_i^x\sigma_{i+1}^x + J_y \sigma_i^y\sigma_{i+1}^y + J_z \sigma_i^z\sigma_{i+1}^z + \;\textrm{boundary terms}.
\ee
Here, $\sigma_i^x$, $\sigma_i^y$ and $\sigma_i^z$ are the usual Pauli matrices acting at site $i$ and various boundary conditions can be specified. For equal coupling parameters $J_x=J_y=J_z=1$, the so-called XXX chain, the Hamiltonian $H$ was diagonalised by Hans Bethe in 1931 by his now famous wave-function Ansatz. Bethe's achievement laid the foundation of a revolution in the 1960's when the field of solvable lattice models came to life. In 2006, the year in which Bethe would have turned 100, JSTAT devoted a topical issue to Bethe's achievement, to celebrate its 75th birthday.  

In a remarkable paper in 2000, Razumov and Stroganov (R-S) \cite{RazuS00} studied the
ground-state wave function of the quantum XXZ Heisenberg chain with periodic boundary conditions at a special value of the coupling parameters ($J_x=J_y=1$, $J_z=-1/2$). They noticed from direct diagonalisation for small numbers of
lattice sites $L=2n+1$, that the largest component and the normalization of the wave function are related to combinatorial objects called ($n\times n$) alternating-sign matrices \cite{Bress99}. Razumov and Stroganov conjectured that these coincidences remain valid for any number of sites, and were thus able to obtain, for \textit{finite} $L$, exact analytical expressions for certain correlation functions. This conjecture triggered a lot of other conjectures which brought together the study of quantum spin chains, conformally invariant stochastic processes and combinatorics.

\subsection*{Loop model}
It was realized in \cite{BatchGN01} that similar conjectures could be made in the well known O($n=1$) loop model, a model for critical percolation clusters on the square lattice. The conjectures were also extended to more general boundary conditions. The connection with the loop model led R-S \cite{RazuS01} to a new refined conjecture (called the R-S conjecture). The R-S conjecture gives a much deeper connection between the ground-state wave function of the XXZ quantum spin chain and alternating sign matrices, or rather fully packed loop models as we have seen above. Conjectures similar to the one of R-S were subsequently also made for loop models and XXZ quantum spin chains with different boundary conditions, relating them to different symmetry classes of ASMs.

The results for the quantum spin chain and the loop model are unified by realizing that they arise from different representations of the same Temperley-Lieb (TL) algebra. The recent paper by Zinn-Justin deals with a special case of the one boundary Temperley-Lieb algebra which arises as a quotient of the Hecke algebra Type B. This algebra has generators $e_i$ ($i=1,\ldots,L$) which obey the following relations,
\begin{eqnarray}
e_i^2 &=& -(q+q^{-1}) e_i,\qquad e_ie_{i\pm 1}e_i =e_i \qquad i=1,\ldots,L-1,\\
e_L^2 &=& e_L.
\end{eqnarray}
The generators $e_i$ are precisely those appearing in (\ref{eq:H_alg}) when $q=\exp(2\i\pi/3)$.

\end{document}